\input{aipcheck}

\documentclass[
    ,final            
  ]
  {aipproc}

\layoutstyle{6x9}

\begin{document}

\title{Are black holes big enough to quench cooling\\ in cluster cool cores?}


\author{Mateusz Ruszkowski}{address={Dept. of Astronomy, The University of Michigan, 500 Church Street, Ann Arbor, MI 48109, USA},
altaddress={The Michigan Center for Theoretical Physics, 3444 Randall Lab, 450 Church St, Ann Arbor, MI 48109, USA} }

\begin{abstract}
Total energy arguments 
(e.g., Fabian et al. 2002)
suggest that black
holes need to have masses significantly in excess of the prediction
from the classic $M_{bh}-\sigma$ relation in order to offset the
cooling losses in massive cool core clusters. This suggests that the
black holes may be too small to "power" such clusters. However, 
Lauer et al. (2007)
argue that the $M_{bh}-L$ relationship is a 
better predictor of black hole masses in high luminosity galaxies
and that this relationship
predicts significantly higher masses in BCGs. They find
slow increase in $\sigma$ with $L$ and a more rapid increase in
effective radii with $L$ seen in BCGs as opposed to less luminous
galaxies. Motivated by these results and the theoretical work of
Boylan-Kolchin et al. (2006) on isolated mergers, we perform high-resolution
cosmological simulations of dry mergers in a massive galaxy cluster
identified in the \emph{Millennium Run} including both the dark matter halos
and stellar bulges of merging galaxies. We demonstrate that the BCG
clearly evolves away from the size-luminosity relation as defined by
the smaller galaxies (i.e., the relation bends) and we also see a
bending in the $L-\sigma$ relation. As $M_{bh}$ is
expected to be proportional to the mass and luminosity of the stellar
bulge of the BCGs (if they were formed in predominantly dissipationless mergers), our
findings are consistent with those of Lauer et al. (2007) on a qualitative
level and suggest that the black holes in BCGs may indeed be more massive than predicted from the 
standard $M-\sigma$ relation.
\end{abstract}

\maketitle

\section{Introduction}

\indent
   The brightest cluster galaxies (BCGs) are special. They are
the most massive and luminous galaxies in the universe. They
are typically located in the very centers of clusters of galaxies
which indicate that their formation is closely linked to that of
the clusters themselves. Their formation history is therefore
distinct from typical elliptical galaxies.
Recent analysis by \cite{Best2007} shows that BCGs
are also more likely to host active galactic nuclei (AGNs) than
other galaxies of the same stellar mass. This indicates that the
supermassive black holes hosted by these  
objects play a pivotal role in quenching cooling flows and star
formation in clusters.

\section{Black hole masses and heating of cool cores}

BCGs have been recently shown to lie off the standard scaling
relations of early-type galaxies (e.g., \cite{Lauer2007}).
In particular, they show excess luminosity (or stellar mass) above
the prediction of the standard Faber-Jackson relation at high
galaxy masses. This hints at the interesting possibility that their
black hole masses may be larger than estimated from the $M-\sigma$
relation. It has been demonstrated 
that the amount of star formation in BCGs is comparatively low
(e.g., O'Dea, these proceedings; Hicks, these proceedings).
Under such conditions mergers between the BCG and other massive ellipticals should 
preserve the ratio between the stellar mass of the merger remnant and the final mass of its central
supermassive black hole. Therefore, any observed departure from the standard
slope of the  $L-\sigma$ relation should translate into bending in the $M-\sigma$ relation.\\
\indent
It has been recognized that simple arguments based on the overall balance of the energy supplied by AGN
and the energy lost to radiative cooling in cool cores over cluster lifetime imply that the masses of the central black holes 
should be larger then expected from the standard $M-\sigma$ relation
(\cite{Fabian2002, Fujita2004}). More specifically, 
assuming standard efficiency of 10\% of conversion of accreted mass to energy, the masses of black holes in the 
hottest clusters would have to exceed $10^{10}M_{\rm sol}$. Alternatively, the efficiencies of these black holes would
have to appreciably exceed 10\%. \\
\indent
Motivated by the observational findings of \cite{Lauer2007} and theoretical work of 
\cite{Boylan-Kolchin2006} on the effect of isolated dry mergers on scaling relations, we performed cosmological 
simulations of the formation of brightest cluster galaxies. The details of the methodology can be found in 
\cite{Ruszkowski2009}. In brief, we identified a very massive cluster in the {\it Millennium Run} and 
traced it back to $z=127$. The initial conditions in the corresponding region were refined beyond the {\it Millennium Run} resolution 
and the simulation was evolved to $z=3$. At this redshift we replaced the most massive halos in a protocluster by galaxy models 
consisting of dark matter and stars. We then evolved the cluster to $z=0$. This approach enabled us to include numerous galaxy 
mergers in a self-consistent fashion while also taking into account the effect of the cluster potential on the properties of the 
forming BCG. 

\section{Results}

Figure 1 (left panel) shows the evolution of the relation between 
the half-light radius and the stellar mass within that
radius. This relation is similar to the Kormendy relation. The
initial ($z=3$) and final ($z=0$) models are represented by
the red and black points, respectively. Intermediate redshifts
are marked by open circles with colors as given in the inset.
The figure demonstrates that, while smaller mass galaxies show
relatively little evolution, the BCG clearly gains a substantial
amount of mass and increases in size. Interestingly, it clearly
evolves off the extrapolation of the initial radius-mass relation.
This effect has been reported in the observational work of \cite{Lauer2007}.
This ``puffing-up'' effect is driven by a large number of mergers that the BCG experiences during the formation process.
Figure 2 (middle panel) demonstrates that the BCG does indeed experience many more mergers than any other galaxy in the cluster.
Consequently, these mergers tend to mix the stars, that are more centrally located, with the dark matter which, in turn, leads to 
the increase in the half-light radius of the BCG. A direct consequence of this mixing process is the increase in the
mass-to-light ratio for the BCG (i.e., the BCG becomes much ``darker'' for a given stellar mass). This is shown in 
the right panel in Figure 1.\\

\begin{figure}
  \includegraphics[height=.24\textheight]{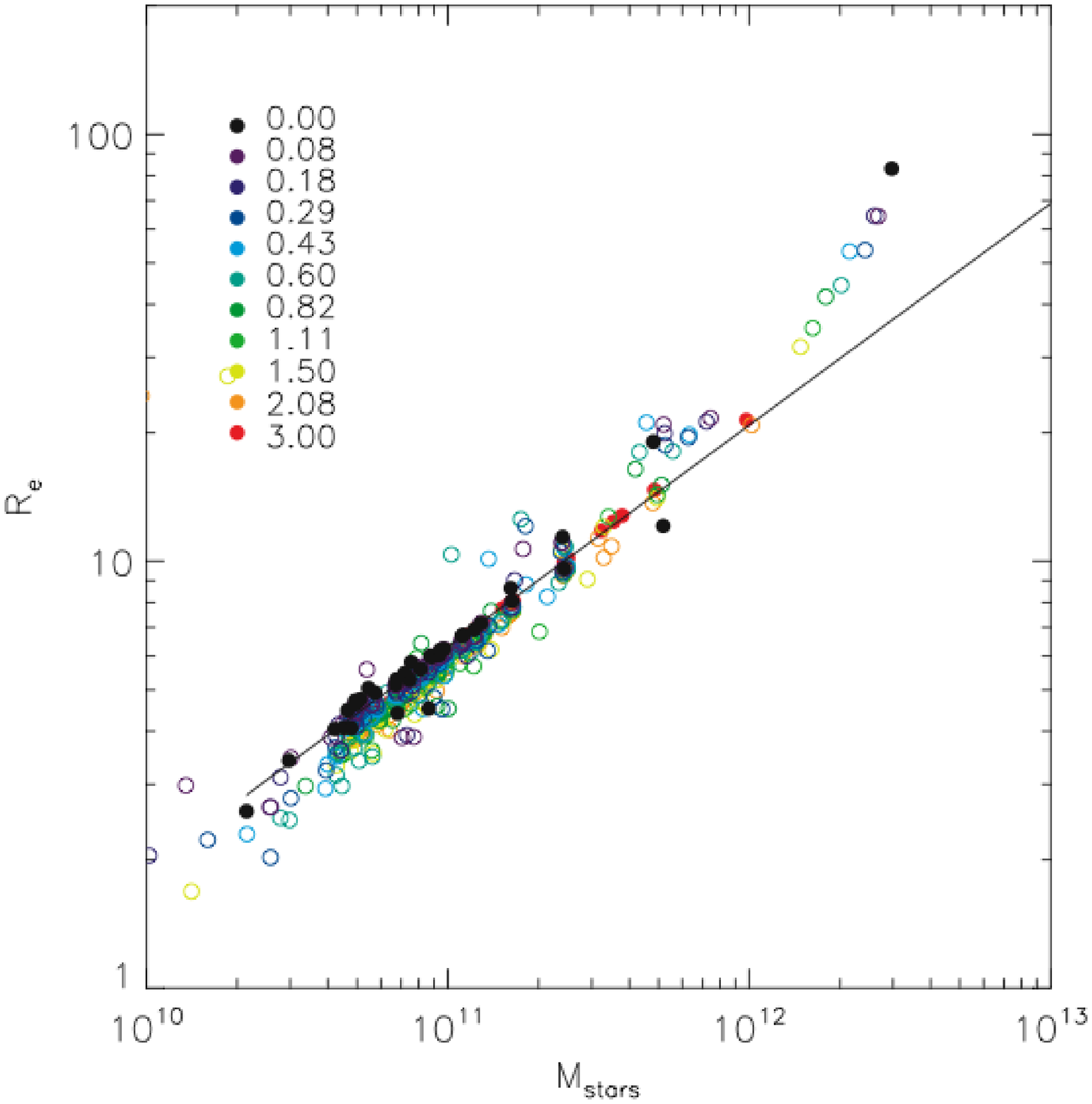}
  \includegraphics[height=.24\textheight]{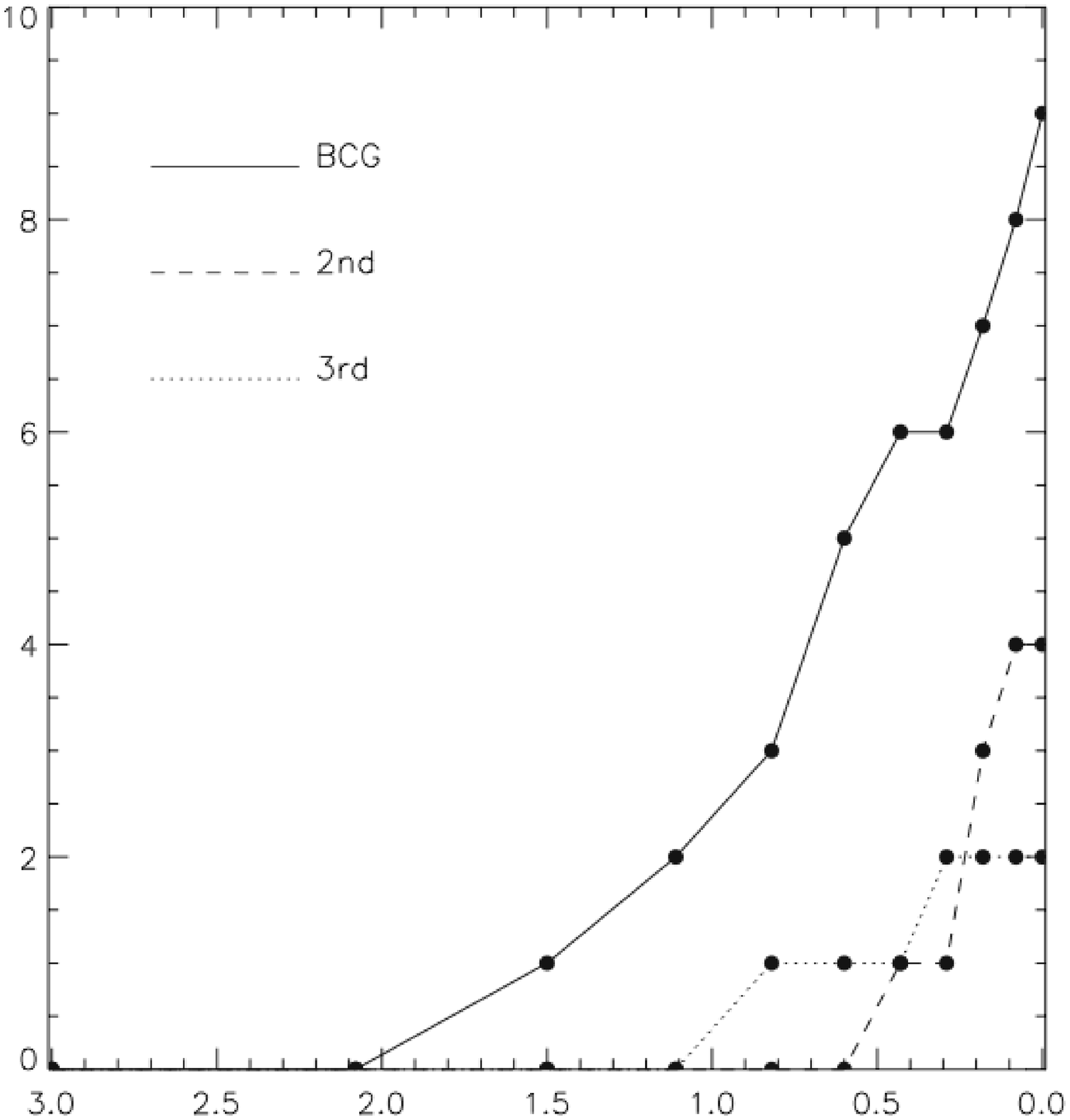}
  \includegraphics[height=.24\textheight]{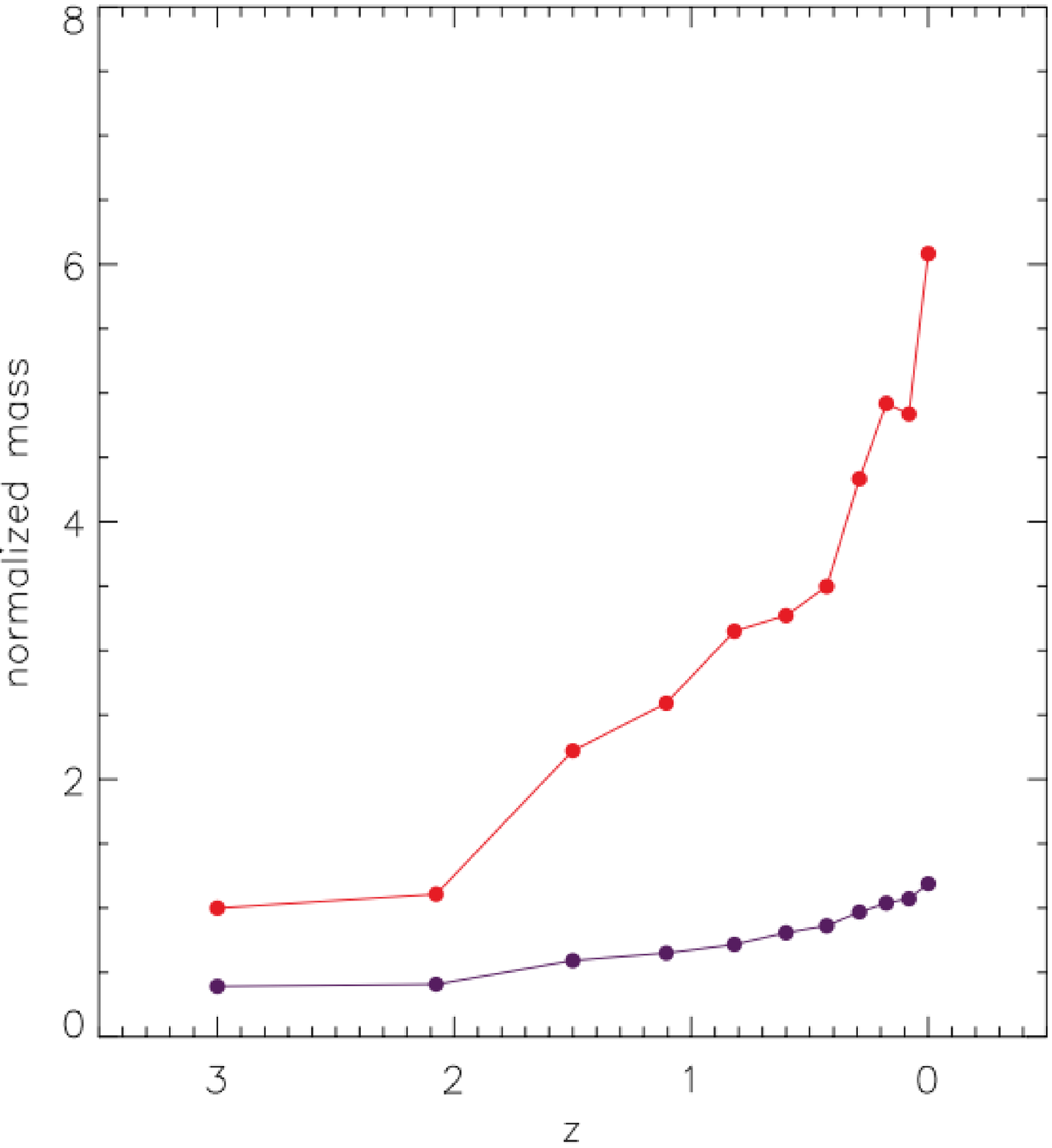}
  \caption{Left panel: Half-light radius vs. stellar mass relation. Filled red circles are for the
           initial models at $z=3$ and the black ones are for $z=0$. Intermediate redshift
           data are shown as open circles and distributed in redshift according to the color
           scale given in the upper left corner.
           Middle panel: cumulative number of mergers experienced by the three most massive galaxies in the cluster as a function of redshift.
           Right panel: Mass in the dark matter (red, top) and stellar (violet, bottom) components for the
           BCG as a function of redshift. }
\end{figure}

\newpage
\indent
{\bf Black hole mass bias in BCGs} 
Merging of the central BCG with cluster galaxies tends to occur on more elongated, lower angular momentum orbits then in the case of 
the mergers between field galaxies. Galaxies on lower angular momentum orbits need to lose less angular momentum to complete the merger.
Consequently, the merger remnant is more extended compared to the size of the remnant resulting from a merger of galaxies on 
high angular momentum orbit (see Figure 2). Now, 
the size of the merger remnant affects its final velocity dispersion. That is, more extended remnants
have relatively lower velocity dispersions. This means that the BCGs, that form as a result of mergers with galaxies on orbits that 
are biased toward low angular momentum (i.e., more ``radial'' mergers), tend to be more extended and, consequently, have somewhat lower 
velocity dispersions then they would have if the mergers were not ``radially-biased''. In other words, the BCG lies somewhat 
above the extrapolated Faber-Jackson relation defined by other (less massive) massive ellipticals in the cluster. Figure 3 shows the ratio 
between the simulated Faber-Jackson relation and its equivalent defined by other ellipticals belonging to the cluster.
This figure clearly demonstrates that the BCG departs from the extrapolated trend. As the final assembly of the BCG is likely 
dominated by dissipationless mergers, any bend in the Faber-Jackson relation can be translated into the bend in the (log-log) 
relation between the mass in the stellar component and its velocity dispersion. If indeed the mergers are predominantly dry, 
then the mass ratio between the stellar component and the central supermassive black hole should be preserved. Since the 
stellar mass of the BCG for a given velocity dispersion is boosted, so should be the masses of the supermassive black holes 
in its center. If so, this mechanism would be one way to reconcile the total energy requirements in cool cores with the 
amount of heating supplied by the central AGN.

\begin{figure}
  \includegraphics[height=.25\textheight]{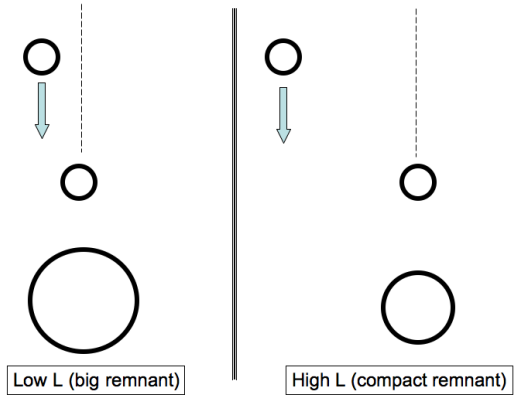}
  \caption{The effect of the orbital parameters of the merging galaxies on the remnant size.}
\end{figure}

\begin{figure}
  \includegraphics[height=.25\textheight]{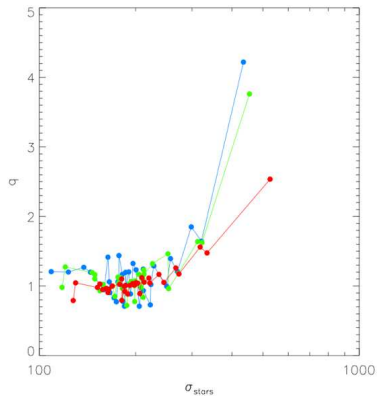}
  \caption{Theoretical boost in the central black hole mass beyond the expectation based on the $M-\sigma$ relation. The curves 
are for different apertures measured in terms of the varying fraction of the half-light radius (1/16, 1/8, 1), from top to bottom. 
For higher apertures around the BCG, the cluster potential has an effect on the inferred velocity dispersions.}
\end{figure}

\begin{theacknowledgments}
M.R. thanks V. Springel (co-author of the paper on which these proceedings are based), 
A. von der Linden, Monica Valluri,
Marta Volonteri, M. Boylan-Kolchin, M. Dotti, O.
Moeller, E. Bell, A. Romanowsky, and S. White for
many stimulating discussions.

\end{theacknowledgments}

\bibliographystyle{aipproc}   


%


\end{document}